\documentclass[iop]{emulateapj}
\usepackage{epsfig}
\usepackage{apjfonts}
\usepackage{amsmath,color,bm}
\usepackage{booktabs}
\usepackage[breaklinks,colorlinks,urlcolor=blue,citecolor=blue,linkcolor=blue]{hyperref}
\shorttitle{Of Harbingers and Higher Modes}
\shortauthors{Kapadia et al.}

\usepackage{natbib}
\usepackage{nameref}
\usepackage{graphicx}
\usepackage{graphics}
\usepackage[space]{grffile}
\usepackage{latexsym}
\usepackage{amsfonts,amsmath,amssymb}
\usepackage{url}
\usepackage[utf8]{inputenc}
\usepackage{fancyref}
\usepackage{hyperref}
\usepackage{xcolor}
\usepackage[normalem]{ulem}

\newcommand{\ffrac}[2]{\ensuremath{\frac{\displaystyle #1}
                                        {\displaystyle #2}}}

\newcommand{\Ylm}{\ensuremath{Y_{\ell m}^{-2}(\iota, \varphi_{o})}}
\newcommand{\blambda}{\pmb{\lambda}}

\begin{document}

\title{Of Harbingers and Higher Modes: \\Improved gravitational-wave early warning of compact binary mergers}


\author{Shasvath J. Kapadia$^1$}
\author{Mukesh Kumar Singh$^1$}
\author{Md Arif Shaikh$^1$}
\author{Deep Chatterjee$^2$}
\author{Parameswaran Ajith$^{1,3}$}
\affiliation{$^1$~International Centre for Theoretical Sciences, Tata Institute of Fundamental Research, Bangalore 560089, India}
\affiliation{$^2$~Department of Physics, University of Wisconsin--Milwaukee, Milwaukee, WI 53211, USA}
\affiliation{$^3$~Canadian Institute for Advanced Research, CIFAR Azrieli Global Scholar, MaRS Centre, West Tower, 661 University Ave, Toronto, ON M5G 1M1, Canada}

\begin{abstract}
A crucial component to maximizing the science gain from the multi-messenger follow-up of gravitational-wave (GW) signals from compact binary mergers is the prompt discovery of the electromagnetic counterpart. Ideally, the GW detection and localization must be reported early enough to allow for telescopes to slew to the location of the GW-event before the onset of the counterpart. However, the time available for early warning is limited by the short duration spent by the dominant ($\ell = m = 2$) mode within the detector's frequency band. Nevertheless, we show that, including higher modes - which enter the detector's sensitivity band well before the dominant mode - in GW searches, can enable us to significantly improve the early warning ability for compact binaries with asymmetric masses (such as neutron-star-black-hole binaries).
We investigate the reduction in the localization sky-area when the $\ell = m = 3$ and $\ell = m = 4$ modes are included in addition to the dominant mode, considering typical slew-times of electromagnetic telescopes ($30-60$ sec).
We find that, in LIGO's projected ``O5’' (“Voyager”) network with five GW detectors, some of the neutron-star-black-hole mergers, located at a distance of $40$ Mpc, can be localized to a few hundred sq. deg. $\sim 45$ sec prior to the merger, corresponding to a reduction-factor of $3-4$ ($5-6$) in sky-area. For a third-generation network, we get gains of up to 1.5 minutes in early warning times for a localization area of $100$ sq. deg., even when the source is placed at $100$ Mpc.
\end{abstract}

\section{Introduction}\label{sec:introduction}
The first gravitational-wave (GW) detection of a binary neutron star merger, GW170817 \citep{GW170817-DETECTION}, also produced an electromagnetic counterpart that was followed up extensively by various telescopes worldwide observing different bands of the electromagnetic spectrum \citep{GW170817-MMA}. This event became a watershed in multi-messenger astronomy, as it demonstrated the immense science gain in observing the same transient in multiple observational windows. GW170817 verified the previously conjectured engine of short gamma-ray bursts (GRBs) as the merger of binary neutron stars (BNS) \citep[see][for a review]{NakarGRB}. In addition, it enabled an unparalleled study of a new class of optical transients called kilonovae \citep{MetzgerKN}, which revealed an important environment in which heavy elements get synthesized \citep{GW170817-HEAVY-ELEMENTS}. The multi-messenger observations also provided stringent constraints on the speed of GWs \citep{GW170817-TGR}, gave important clues to the nuclear equation of state at high densities \citep{GW170817-SOURCE-PROPERTIES, GW170817-EOS}, and an independent estimation of the Hubble constant~\citep{GW170817-HUBBLE}.

An early warning of the merger from the GW data would allow many additional science benefits: For example, it would enable the observations of possible precursors~\citep{2012PhRvL.108a1102T}, a better understanding of the kilonova physics and the formation of heavy elements by identifying the peak of kilonova lightcurves~\citep{Drout2017,Cowperthwaite2017}, and possible signatures of any intermediate merger product (e.g., \cite{HotokezakaHNS}) that might have been formed.

BNS mergers are traditionally expected to produce EM counterparts, and therefore it is not surprising that the first efforts towards GW early-warning focused on such events. The inspiral of BNSs lasts for several minutes within the frequency band of ground based GW detectors. If sufficient signal-to-noise ratio (SNR) could be accumulated during this time, ideally tens of seconds to a minute before merger, it could allow for a tight enough sky map for telescopes, enabling them to point at the binary before it merges \citep{CannonEW}.

Early warning for heavier binaries, like neutron-star-black-hole (NSBH) binaries or binary black holes (BBHs), is more challenging, given that they spend significantly smaller durations in the band of ground based detectors (for e.g., GW150914 spent $\sim0.1$s in the LIGO detectors' frequency band \citep{GW150914-DETECTION}) \footnote{Stellar mass BBH mergers are not expected to produce electromagnetic emissions under standard scenarios. Nevertheless, there are proposals of possible counterparts to such mergers (e.g., \cite{Loeb_2016}). The Fermi satellite had also announced a candidate gamma-ray counterpart coincident with GW150914 \citep{Connaughton_2016}; additionally, the Zwicky Transient Facility recently announced a candidate optical counterpart to a candidate BBH event \citep{ZTF-BBH}.}. A possible way to achieve early warning is to detect these systems early in the inspiral, although that would require ground-based detectors to be sensitive at very low frequencies. Seismic noise being the dominant impediment to such low-frequency detections, a ``multi-band'' detection strategy has been proposed, where the upcoming space-based detector LISA would detect the binary early in its inspiral, potentially years before it reaches the frequency band of ground based detectors \citep{SesanaMultiband}.

In this \emph{letter}, we describe an alternative method for early-warning targeted at unequal-mass compact binaries (especially NSBHs), that could be applied to the upcoming second and third generation (2G and 3G) network of ground-based detectors \citep{LIGOProspects, CE, ET}. The method essentially relies on the fact that the detected GW signal from asymmetric binary inspirals, within a range of inclination angles, could contain contributions from several higher modes in addition to the dominant, quadrupole ($\ell = m = 2$) mode~\citep[e.g.,][]{Varma:2014jxa}. Since the majority of the higher modes (with $m > 2$) oscillate at larger multiples of the orbital frequency than the dominant mode, we expect these higher modes to enter the frequency band of the detector well before the dominant mode. Thus, using GW templates including higher modes in online GW searches \citep[e.g.,][]{gstlal, mbta, pycbc, spiir} would enable us to detect and localize the binary earlier than analyses that only use the dominant mode, potentially allowing significant reduction in localisation sky-areas, for early-warning times that are comparable to the slew times of a number of electromagnetic telescopes ($\sim 30-60$ seconds).  

We investigate the reduction in the sky-area (as compared to the same using only the dominant mode) by including the higher modes $\ell = m = 3$ and $\ell = m = 4$ in addition to the dominant mode. We consider binaries with secondary masses spanning the range $m_2 = 1-2.5 M_{\odot}$, mass-ratios $q := m_1/m_2 = 4-20$, located at a GW170817-like distance ($d_L \simeq 40$ Mpc). We find that, for a network of five detectors with projected sensitivities pertaining to the 5th observing run (O5) \citep{observer_summary}, we get 
a reduction of sky-area from a few thousand to a few hundred square degrees (a factor of $3-4$), for an early warning time of 45 seconds. These gains increase to 
factors of $5-6$ for the same detector-network with the three LIGO detectors, including LIGO-India~\citep{LIGO-INDIA}, upgraded to ``Voyager'' sensitivity~\citep{Adhikari:2019zpy}. 
In a 3G network consisting of two Cosmic Explorers and one Einstein Telescope, for a localisation of 100 sq. deg., we can get early-warning-time gains of up to $1.5$ minutes, for binaries located at 100 Mpc. 

The \emph{letter} is organized as follows. Section \ref{sec:ew_hm} elaborates on the early-warning method, while also describing higher modes and giving quantitative arguments as to why one should expect them to enhance early-warning times. Section \ref{sec:results} describes the results, in particular the reduction in localisation area of the sources by including higher modes and the effect of extrinsic parameters on this reduction, for various upcoming observing scenarios involving ground-based GW detectors. Section \ref{sec:conclusion} gives a summary and an assessment of the benefits of the proposed method.

\section{Early Warning with Higher Modes}\label{sec:ew_hm}
The gravitational waveform, conveniently expressed as a complex combination $h(t) := h_{+}(t) - ih_{\times}(t)$ of two polarizations $h_{+}(t)$ and $h_{\times}(t)$, can be expanded in the basis of spin $-2$ weighted spherical harmonics $\Ylm$ \citep{NewmanPenrose} 
\begin{equation}
h(t; \iota, \varphi_{o}) = \ffrac{1}{d_{L}} \sum_{\ell = 2}^{\infty}\sum_{m = -\ell}^{\ell} {h_{\ell m}(t, \blambda)} \, \Ylm.
\end{equation}
Here, $d_L$ is the luminosity distance, and $h_{\ell m}$ are the multipoles of the waveform that depend exclusively on the intrinsic parameters of the system $\blambda$ (component masses, spins, etc.) and time $t$. On the other hand, the dependence of the waveform on the orientation of the source with respect to the line-of-sight of the detector is captured by the basis functions $\Ylm$ of the spin $-2$ weighted spherical harmonics, where $\iota, \varphi_{o}$ are the polar and azimuthal angles in the source-centered frame, that define the line of sight of the observer with respect to the total angular momentum of the binary. 
For non-precessing binaries, due to symmetry, modes with negative $m$ are related to the ones with corresponding positive $m$ by $h_{\ell-m} = (-1)^\ell h_{\ell m}^*$. In this \emph{letter}, we consider only non-precessing binaries. Hence, even when we mention only modes with positive $m$, it is implied that the corresponding $-m$ modes are also considered.  

The dominant multipole corresponds to $\ell= m =2$, which is the quadrupole mode. The next two subdominant multipoles are $\ell= m = 3$ and $\ell= m = 4$. The contribution of subdominant modes relative to the quadrupole mode depends on the asymmetries of the system --- for e.g., relative contribution of higher modes is larger for binaries with large mass ratios. Also, due to the nature of the spin $-2$ weighted spherical harmonics, the higher-mode contribution to the observed signal is the largest for binaries with large inclination angles (say, $\iota = 60$ deg.). 

\begin{figure}[t]
  \begin{center}
    \includegraphics[width=1.0\columnwidth]{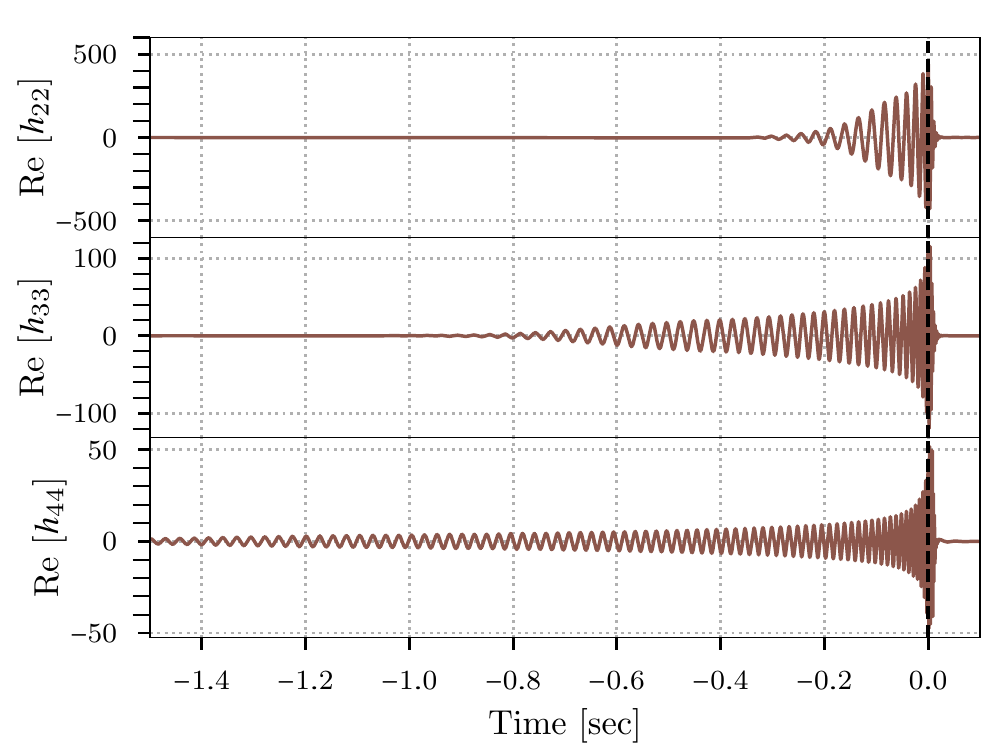}
  \end{center}
  \caption{Schematic illustration of how different modes appear in the detector band. We show the real part of the \emph{whitened} modes $h_{\ell m}$ (with $\ell = m = \{2, 3, 4\}$) of a compact binary coalescence waveform ($d_L = 500$ Mpc, $q = 5$, $m_1 + m_2 = 80 M_{\odot}$), as a function of time. The modes are whitened by the noise PSD of Advanced LIGO to show their expected contribution to the SNR. The higher the $m$, the earlier it enters the frequency band of the detector. This can be seen by the appearance of the non-zero amplitudes of the higher modes at a time $\Delta \tau$ before the merger (dashed black vertical line), where $\Delta \tau$ increases with increasing $m$.}
  \label{fig:hm_illustrative}
\end{figure}

The instantaneous frequency of each spherical harmonic mode is related to the orbital frequency in the following way (assuming a non-precessing orbit):
\begin{equation}
F_{\ell m}(t) \simeq m \, F_{\mathrm{orb}}(t).
\end{equation}
Thus, higher modes (with $m > 2$) enter the frequency band of the detector (say, 10 Hz) before the dominant mode (see Fig.~\ref{fig:hm_illustrative} for a qualitative illustration) .
The time taken by the binary to merge, once it has reached an orbital frequency of $F_{\mathrm{orb}}$, is approximately given by~\citet{Sathyaprakash:1994nj}
\begin{equation}
\tau \simeq \frac{5}{256}\,\mathcal{M}^{-5/3}\, (2\pi F_\mathrm{orb})^{-8/3} \propto (F_{\ell m}/m)^{-8/3},
\label{eq:chirptime}
\end{equation}
where $\mathcal{M} := (m_1 m_2)^{3/5}/(m_1 + m_2)^{1/5}$ is the chirp mass of the binary. Thus, the in-band duration of a higher mode $h_{\ell m}$ is a factor $(m/2)^{8/3}$ larger than the corresponding $\ell = m = 2$ mode. For the $\ell = m = 3$ mode, this amounts to $\sim 3$ fold increase in the observable duration as compared to the $\ell = m = 2$ mode, and for the $\ell = m = 4$ mode a $\sim 6$ fold increase 

However, the time gained in reaching a fiducial threshold-SNR or localizing a source to a fiducial sky-area, and the reduction in sky-area at a given early-warning time,  depend on two competing factors: On the one hand, higher modes are excited only for binaries with large mass ratios (and hence larger chirp masses, when we fix a lower limit on $m_2 \simeq 1 M_\odot$). On the other hand, according to Eq.\eqref{eq:chirptime}, heavier binaries will merge quicker in the detector band. Thus, the region of the $m_1-m_2$ plane that maximizes the time gains corresponds to regions where the masses are sufficiently asymmetric to excite the higher modes significantly, while not too heavy to make the system hurry through the frequency band of the detector; this region will change depending on the sensitivities of the detectors. Similarly, the reduction in sky-area at a given early-warning time depends additionally on the choice of early-warning time. 

For stationary Gaussian noise (which we assume throughout this paper), the power-spectral-density (PSD) of the noise completely determines its statistical properties. Based on this assumption, assessing a trigger to be worthy of follow-up can be reduced to setting a threshold on the SNR, corresponding to a given false alarm probability. The localization area, at a given confidence, is determined to a good approximation by the separation of the detectors, their individual effective bandwidths, and the SNRs. 
We use the method proposed by \citet{Fairhurst1, Fairhurst2} to estimate the sky area from the times of arrival of the signal at the detectors, and timing uncertainties. In this method, the localization sky area of a source located at a given sky location can be computed from the pair-wise separation of the detectors, as well as each detectors' timing errors. Note that if the detectors are approximately co-planar, then the mirror degeneracy with respect to the plane of the detector needs to be broken by additional waveform consistency tests between detectors. 

\section{Results}\label{sec:results}
\begin{figure*}[t]
  \begin{center}
    \includegraphics[width=2\columnwidth]{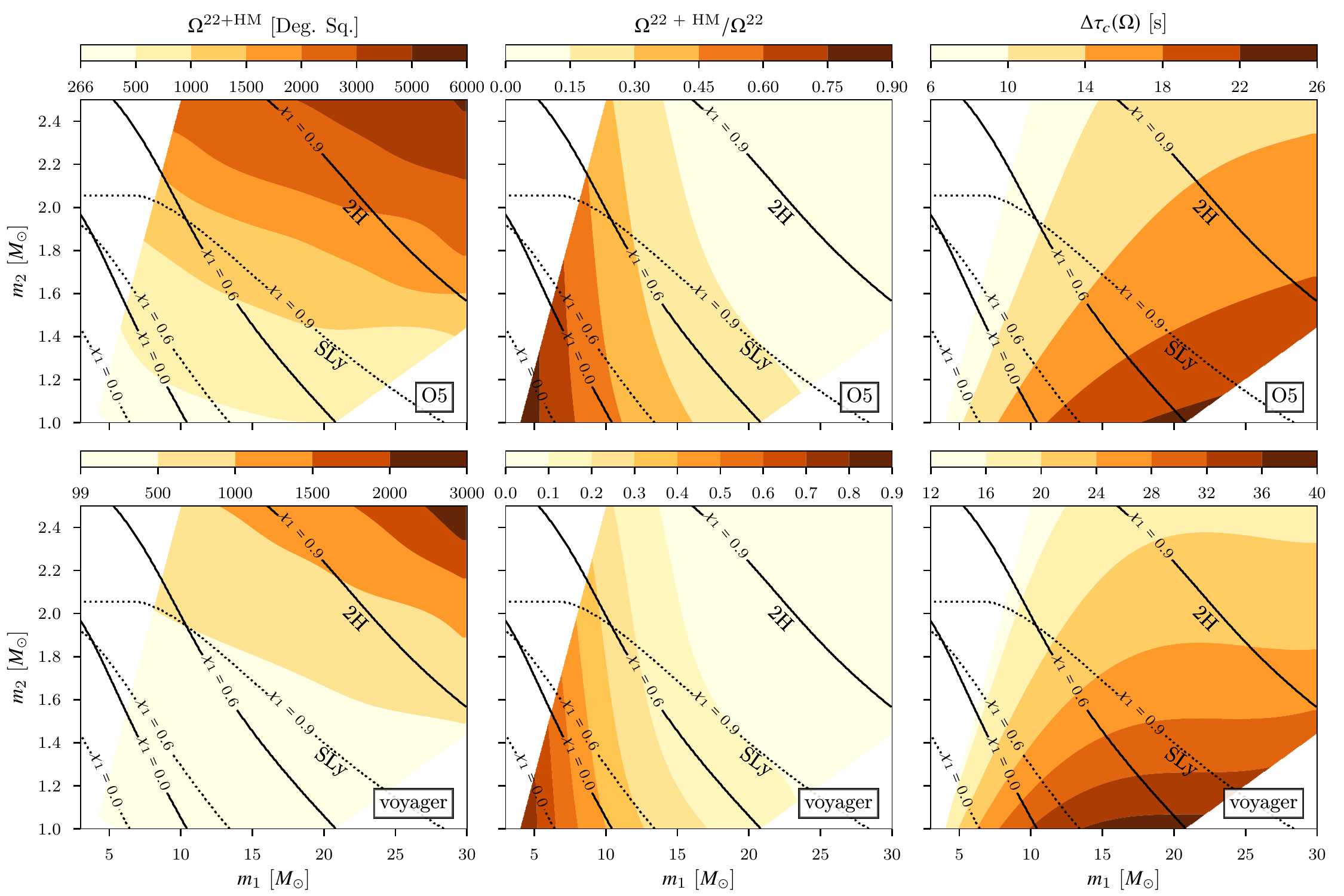} 
  \end{center}
  \caption{Left plots show the localization sky-area (at $90\%$ confidence) using higher modes, for an early warning time of 45 sec. Middle plots show the same as a fraction of the sky area achieved using only the dominant modes. Right plots show the gains in the early warning time for a fiducial sky-area of 1000 sq. deg, due to the inclusion of higher modes. These plots correspond to binaries with $m_2 = 1-2.5 M_{\odot}$ and $q = 4-20$, located at 40 Mpc (other extrinsic parameters set to their optimal values, with inclination $\iota = 60$ deg.). The compact objects are assumed to be non-spinning; however, even including a primary spin as large as $\chi_1 = 0.9$ does not alter these results significantly. We also plot the contours that demarcate the region corresponding to binaries that would produce a non-zero ejecta mass suggesting the possibility of EM counterpart, for various $\chi_1$ values~\citep{FoucartEMB}. Two sets of contours, corresponding to 2H (solid, black contours) and SLy (dotted, black contours) EOSs are plotted. When higher modes are employed, early-warning sky localization of few hundred square degrees can be achieved over significant fraction of the putative ``EM-bright'' region of the parameter space. This is a factor of $3-4$ ($5-6$) reduction in sky area as compared to the same using only the dominant mode in O5 (Voyager). The early warning time gains can be as much as $\sim 25 \, (40)$ secs in the O5 (Voyager) scenario. 
  }
  \label{fig:results-delta-t-em-bright}
\end{figure*}

\begin{figure}[t]
  \begin{center}
    \includegraphics[width=1.0\columnwidth]{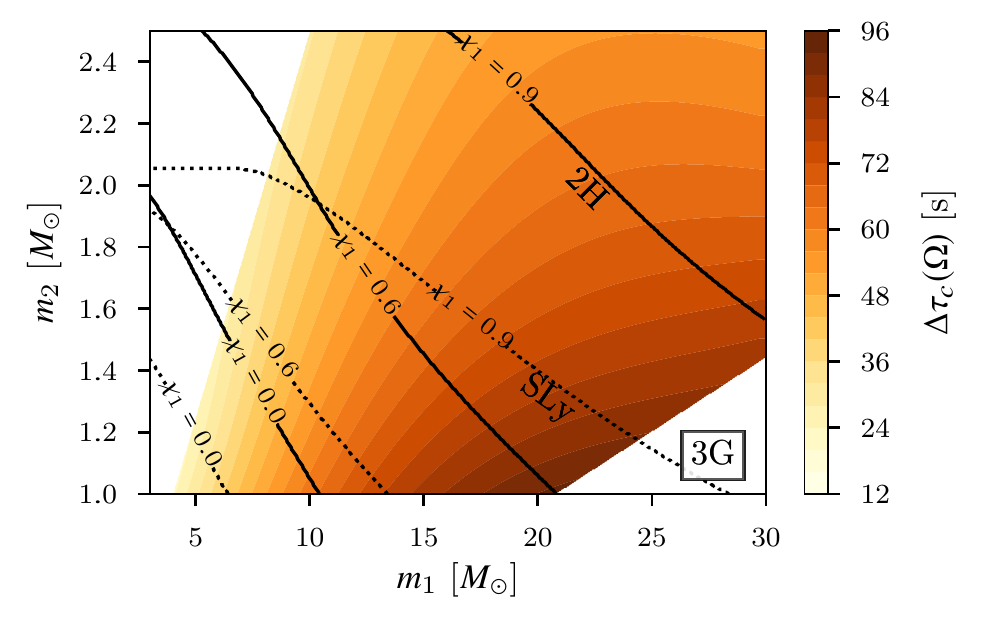} 
  \end{center}
  \caption{Gains in early-warning time upon inclusion of higher modes, for the 3G scenario, assuming a fiducial sky-area of 100 sq. deg., and sources located at 100 Mpc. The other extrinsic parameters are set to their optimal values, with inclination $\iota = 60$ deg. These gains can be as much as 1.5 minutes for relatively low-mass systems that are highly asymmetric. For binaries that are likely to have EM counterparts even for moderate to low spins of the primary mass, the gains can be as much as a minute. 
  }
  \label{fig:results-delta-t-em-bright-3G}
\end{figure}

We generate two sets of GW signals --- one set containing just the $\ell = m = 2$ mode, while the other includes the $\ell = m = 3$ and $\ell = m = 4$ modes in addition to the dominant mode. These are generated using the \textsc{IMRPhenomHM} model \citep{imrphenomhm}, as implemented in the \textsc{LALSuite} software package \citep{lalsuite}. We consider three observing scenarios. The first is the ``O5'' scenario, consisting of LIGO-Hanford, LIGO-Livingston, Virgo, KAGRA and LIGO-India. We assume the most optimistic projected sensitivities, from the document \cite{observer_summary}. Since KAGRA's projected sensitivity for O5 only has a lower limit, we assume that KAGRA's sensitivity will equal Virgo's. The second is the ``Voyager'' scenario \citep[noise PSD taken from][]{Voyager_PSD}, where we assume that all three LIGO detectors, including LIGO-India, will be upgraded to Voyager sensitivity, while Virgo and KAGRA will operate at their O5 sensitivities. The third is the 3G scenario, where the assumed network consists of two Cosmic Explorer detectors, and one Einstein Telescope. {The projected PSD for the Einstein Telescope is taken from \cite{ET_PSD}, and that for Cosmic Explorer is taken from \cite{CE_PSD}}.

In Fig.~\ref{fig:results-delta-t-em-bright}, we summarize the reduction in localization sky-area for a fiducial early warning time of 45 seconds, comparable to the slew times of a number of electromagnetic telescopes ($\sim 30-60$ seconds), and the early-warning time gained by the inclusion of higher modes for a fiducial sky-area of 1000 sq. deg. We focus on unequal-mass binary systems located at a distance of $40$ Mpc \citep{GW170817-DETECTION}, with the secondary mass spanning {$m_2 = 1-2.5 M_{\odot}$} and the mass ratio spanning {$q = 4-20$}~\footnote{We do not consider mass ratios $q > 20$ because the waveforms that we use are calibrated to numerical relativity results only for binaries with $q \lesssim 20$ \citep{imrphenomhm}. We also do not show the results for $q < 4$ (and $m_2 > 2.5 M_{\odot}$), as the improvements are not significant.}. 

We only show results for the case of non-spinning binaries, since our results do not change appreciably with spin. Furthermore, we focus on the mass range that corresponds to NSBHs. This is done for two reasons: the first is that an EM counterpart for binaries detectable by ground based detectors are expected to require a NS; the second is that higher modes are excited predominantly for binaries with asymmetric masses ($q \gg 1$). 
We also include contours that demarcate the region of the $m_1-m_2$ plane that are expected to produce EM-counterparts, based on the spin of the primary and the NS equation of state (EOS) of the secondary \citep{FoucartEMB}~\footnote{Note that these regions correspond to the parameter space of binaries that are expected to produce non-zero dynamical ejecta, which is a necessary, but not sufficient, condition for producing EM counterparts.}. For this purpose, we consider three values of the spin ($0, 0.6, 0.9$), and two EOS: 2H \citep{2H} and SLy \citep{SLy}. The former is a ``stiff'' EOS, predicting a relatively broad region of the component-mass space to produce EM counterparts, while the latter is a more ``realistic'' EOS, as indicated by the GW-based investigations of the properties of GW170817~\citep{GW170817-SOURCE-PROPERTIES}. 

We find that, when higher modes are employed, early-warning sky localization of a few hundred sq. deg. can be achieved over a significant fraction of the putative ``EM-bright'' region of the parameter space (Fig.~\ref{fig:results-delta-t-em-bright}). This corresponds to a factor of $3-4$ ($5-6$) reduction in sky area as compared to the same using only the dominant mode in O5 (Voyager). The early warning time gains can be as much as $\sim 25 \, (40)$ secs in the O5 (Voyager) scenario, for a fiducial sky-area target of 1000 sq. deg. Additionally, Fig.~\ref{fig:results-delta-t-em-bright-3G} summarizes the early-warning time gains in reaching a localization area of 100 sq. deg., for systems located at 100 Mpc, in the 3G scenario. For binaries that are likely to have EM counterparts even for moderate to low spins of the primary mass, the gains can be as much as a minute.

\begin{figure}[t]
    \begin{center}
    \includegraphics[width=1.0\columnwidth]
        {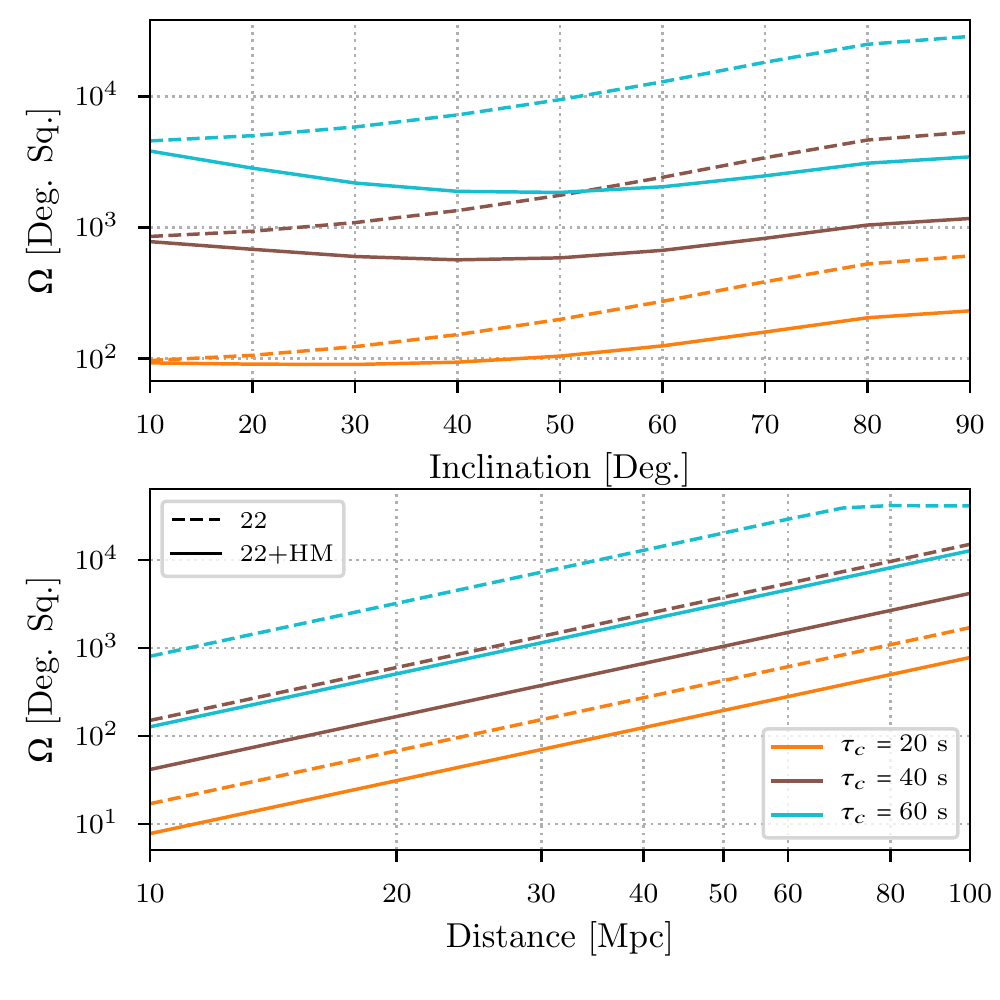}
    \end{center}
    \caption{Variation of the sky-area with inclination angle (top panel) and distance (bottom panel) with and without the inclusion of higher modes, for three early-warning times, in the O5 scenario. We pin the masses to $m_1 = 15M_{\odot}, m_2 = 1.5M_{\odot}$, distance to $40$ Mpc, inclination to $\iota = 60$ deg. (for the bottom panel), and  all other extrinsic parameters to their optimal values. The sky-area (especially when higher modes are included) is only weakly dependent on the inclination angle, while the improvement over the dominant mode is typically the largest for higher inclination angles (where the relative contribution of higher modes is the largest). Sky-area scales as the square of the distance, as expected.} 
    \label{fig:results-skyarea-delta-t-var-dist}
\end{figure}

While Figs.~\ref{fig:results-delta-t-em-bright} and \ref{fig:results-delta-t-em-bright-3G} use fixed values of the location and orientation of the binary, Fig.~\ref{fig:results-skyarea-delta-t-var-dist} shows the variation of the early-warning sky-area as a function of the inclination angle (top plots) and distance (bottom plots), for a binary with $m_1 = 15 M_{\odot}, m_2 = 1.5 M_{\odot}$~\footnote{This choice of masses does not correspond to the optimal mass-combination for time gain, within the mass-space we consider. Nevertheless, it does represent a system that could potentially have an EM counterpart for a moderately $\chi_1 \sim 0.6$ spinning primary assuming a 2H equation of state.}. The sky-area (especially when higher modes are included) is only weakly dependent on the inclination angle, while the improvement over the dominant mode is typically the largest for higher inclination angles (where the relative contribution of higher modes is the largest). In contrast, sky area scales as the square of the distance, as expected. 

\begin{figure}[t]
    \begin{center}
    \includegraphics[width=1.0\columnwidth]
        {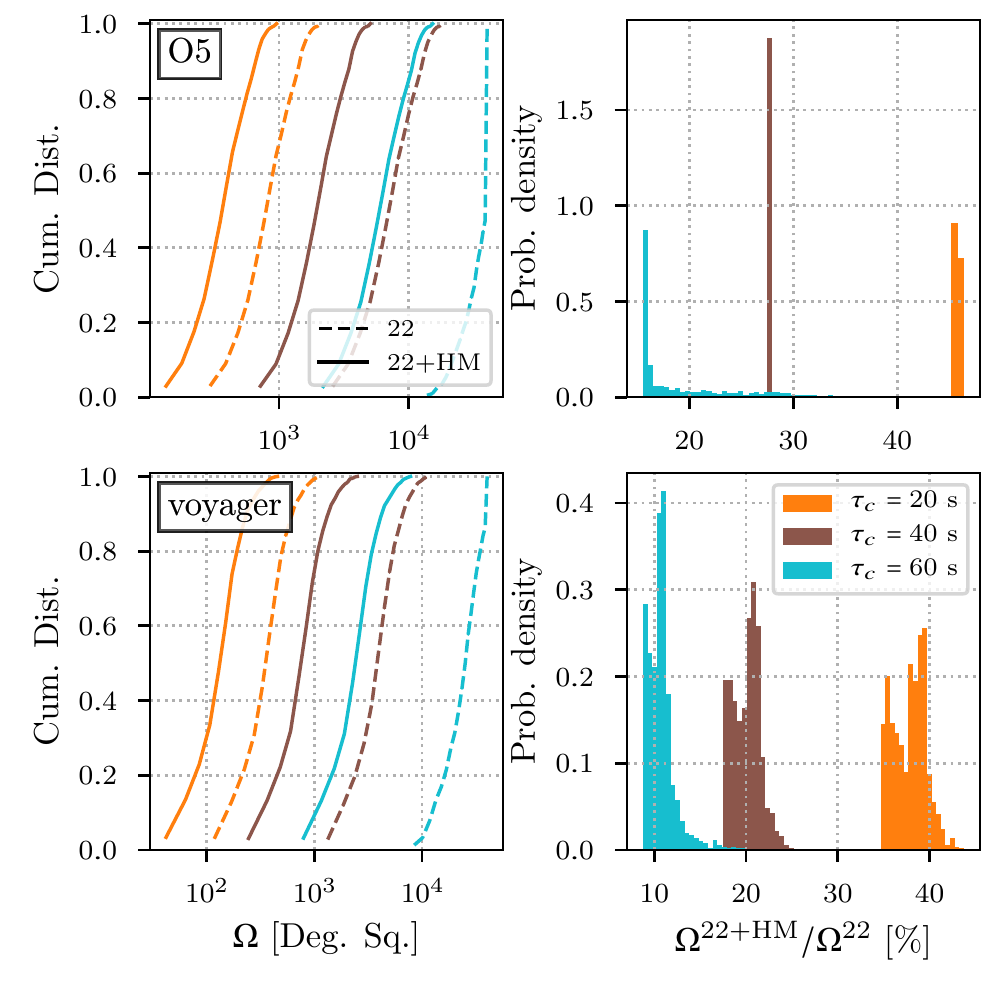}
    \end{center}
    \caption{Distributions of the sky-areas for early-warning times of $20, 40, 60$ secs, with (solid lines) and without (dashed lines) the inclusion of higher modes. The distributions are computed by sampling the parameter space of RA, Dec and polarization, for a $m_1 = 15 M_{\odot}, m_2 = 1.5 M_{\odot}$ system. We set the inclination to $\iota = 60$ deg., and the distance to $40$ Mpc. The rows correspond to different observing scenarios and the columns correspond to (left) sky-area in sq. deg, and (right) the ratio of the sky-areas with/without higher modes. The median sky-area after the inclusion of the higher modes are 500, 2000 and 8000 (200, 800 and 2000) sq. deg, respectively, for the O5 (Voyager) scenario. These correspond to sky-area ratios of $15\%$, $28\%$ and $45\%$ ($10\%$, $20\%$ and $40\%$) for O5 (Voyager).}
    \label{fig:results-skyarea-percent-hist}
\end{figure}

Fig. \ref{fig:results-skyarea-percent-hist} shows the distribution of the early-warning sky-area for the same binary ($m_1 = 15 M_{\odot}, m_2 = 1.5 M_{\odot}$) after fixing the inclination to $60$ deg and distance to 40 Mpc, while randomizing over the sky location and polarization angles. For early-warning times of $20, 40, 60$ secs, the median sky-area after the inclusion of the higher modes are 500, 2000 and 8000 (200, 800 and 2000) sq. deg, respectively, for the O5 (Voyager) scenario. We find that for early-warning times of $20$ and $40$ secs, the reduction factor distributions in sky-area are sharply peaked around $28\%$ and $47\%$ ($20\%$ and $40\%$) respectively, for O5 (Voyager).  This suggests that the reduction factor is mostly independent of sky location. For a $60$ sec early warning time, the reduction factor distributions have longer tails, which is a consequence of the fact that for certain sky locations, the sky-area using just the dominant mode saturates to its maximum possible value ($\sim 40,000$ sq. deg).

We also estimate the time gained in reaching a SNR threshold of 4 for trigger-selection with the inclusion of higher modes. For O5 (Voyager), we get gains of up to $\sim 1$ (2) min, which correspond to a gain of up to $\sim 50\%$ ($80\%$), as compared to the same using dominant mode. For 3G, the gains reach $50$ minutes, corresponding to a $500 \%$ increase. This could potentially be useful for wide-field telescopes (e.g. all-sky gamma-ray burst monitors) to discover precursors and prompt emission and to trigger follow-up observations, as we discuss in the next section.
 
\section{Summary and Outlook}\label{sec:conclusion}
Early localization of compact binary mergers from GW data is useful for fast and wide-field surveys (e.g., \cite{Bellm_2018, Ivezic_2019}) to begin follow-up observations (slew times $\sim 30-60$~s). However, we are limited by the short duration spent by the dominant mode of the gravitational radiation in the detector band. In this \emph{letter}, we showed that the inclusion of higher modes in GW low-latency searches will significantly improve the GW early-warning abilities of GW telescopes \footnote{We note that designing sensitive and computationally tractable searches using waveform templates including the effect of higher modes is not a fully solved problem, although efforts in this direction are ongoing \cite[see, e.g.][for some recent work]{Harry:2017weg}}. This is especially true for asymmetric mass compact binaries with inclined orbits, where higher multipoles of the gravitational radiation are expected to make appreciable contributions to the signal. Recent GW observations have confidently established the existence of some of these higher modes~\citep{GW190412, GW190814}.
 
We find that, using the upcoming five-detector network (O5 or Voyager-type sensitivities) some of the neutron-star-black-hole mergers, located at a distance of $40$ Mpc, can be localized to a few hundred square degrees $\sim 45$ sec prior to the merger (Fig.~\ref{fig:results-delta-t-em-bright}). This corresponds to a factor of $3-4$ ($5-6$)  reduction in sky area in the O5 (Voyager) scenario.  For a third-generation network, we get gains of up to $1.5$ minutes in early warning times for a localization area of 100 sq. deg., even when the source is placed at $100$ Mpc (Fig.~\ref{fig:results-delta-t-em-bright-3G}).

Early-warning localization of a few hundred square degrees is still too large for a single optical telescope to observe in a small number of pointings. However, coordinated observations of several optical telescopes, assisted by a galaxy catalog, can probe the estimated localization region \citep[see, e.g.,][]{SlewingStrategy}. Additionally, such early warning times could allow telescopes to start slewing towards the general location of the event, thus saving up to $\sim$ a minute if the telescopes are pointing in very different directions from the source location. With sufficient coordination between the GW network and automated telescopes, it should be possible to track the shrinking localization area, thus optimizing follow-up. 

Early warning of a merger might allow wide-field gamma-ray telescopes (with field of view of thousands of sq. deg) to trigger on precursors and weak prompt emissions using online sub-threshold analyses, potentially enabling rapid localization of the merger to sub-degree precision~\citep{Swift}~\footnote{Most of the short gamma-ray bursts are expected to have jet opening angles $\lesssim 20$ deg~\citep{Berger:2013jza}, for which the expected improvements in the early warning time due to higher modes are modest (Fig.~\ref{fig:results-skyarea-delta-t-var-dist}). Note, however, that the gamma-ray burst associated with GW170817 was observed as off-axis, with estimated inclination angle equivalent to $\iota \sim 30$ deg~\citep{GW170817-SOURCE-PROPERTIES}.}. This can facilitate a hierarchy of followup observations using x-ray, ultraviolet and optical telescopes with much smaller fields of view. In the absence of GW early warning, such subthreshold triggering algorithms might not be used to identify the gamma-ray counterpart, and thus the followup observations would not be triggered~\citep{FermiGBM}.

Advanced early warning with sky areas spanning thousands of square degrees could be also useful for wide-field radio telescope arrays to dump all the relevant data into storage. This raw data can be analyzed offline to scan the entire localization area to discover any precursors or prompt emission from the merger. 

One might also ask how often would we expect to see NSBH mergers with counterparts, within $\sim 100$ Mpc? Using upper limits on the rate of such mergers from LIGO-Virgo data \citep{GWTC-1}, as well as from models of formation channels, we could do a more elaborate population study to estimate a distribution of early-warning time gains. We are currently in the process of producing these results, which we hope to report soon. Even if the rates are not very high, a small number of such golden events might offer some unique glimpses to the complex physics of compact binary mergers.

\paragraph{Acknowledgments:} 
We are grateful to Shaon Ghosh for reviewing our manuscript and providing useful comments. We also thank Stephen Fairhurst for clarifications on \citep{Fairhurst1, Fairhurst2}, Srashti Goyal for help with the \textsc{PyCBC} \citep{pycbc_software} implementation of the antenna pattern functions, and Varun Bhalerao and Shabnam Iyyani for suggestions on possible uses of early warning for gamma-ray and radio telescopes. SJK's, MKS's, MAS's and PA's research was supported by the Department of Atomic Energy, Government of India. In addition, SJK's research was funded by the Simons Foundation through a Targeted Grant to the International Centre for Theoretical Sciences, Tata Institute of Fundamental Research (ICTS-TIFR). PA's research was funded by the Max Planck Society through a Max Planck Partner Group at ICTS-TIFR and by the Canadian Institute for Advanced Research through the CIFAR Azrieli Global Scholars program. DC would like to thank the ICTS-TIFR for their generous hospitality; a part of this work was done during his visit to ICTS.

\begin{thebibliography}{}
\expandafter\ifx\csname natexlab\endcsname\relax\def\natexlab#1{#1}\fi
\providecommand{\url}[1]{\href{#1}{#1}}
\providecommand{\dodoi}[1]{doi:~\href{http://doi.org/#1}{\nolinkurl{#1}}}
\providecommand{\doeprint}[1]{\href{http://ascl.net/#1}{\nolinkurl{http://ascl.net/#1}}}
\providecommand{\doarXiv}[1]{\href{https://arxiv.org/abs/#1}{\nolinkurl{https://arxiv.org/abs/#1}}}

\bibitem[{{Abbott} {et~al.}(2018{\natexlab{a}}){Abbott}, {Abbott}, {Abbott},
  {Acernese}, {LIGO Scientific Collaboration}, \& {Virgo
  Collaboration}}]{GW170817-EOS}
{Abbott}, B.~P., {Abbott}, R., {Abbott}, T.~D., {et~al.} 2018{\natexlab{a}},
  \prl, 121, 161101, \dodoi{10.1103/PhysRevLett.121.161101}

\bibitem[{{Abbott} {et~al.}(2018{\natexlab{b}}){Abbott}, {Abbott}, {Abbott},
  {et~al.}}]{LIGOProspects}
---. 2018{\natexlab{b}}, Living Reviews in Relativity, 21, 3,
  \dodoi{10.1007/s41114-018-0012-9}

\bibitem[{Abbott {et~al.}(2016)}]{GW150914-DETECTION}
Abbott, B.~P., {et~al.} 2016, Phys. Rev. Lett., 116, 061102,
  \dodoi{10.1103/PhysRevLett.116.061102}

\bibitem[{Abbott {et~al.}(2017{\natexlab{a}})}]{GW170817-DETECTION}
---. 2017{\natexlab{a}}, Phys. Rev. Lett., 119, 161101,
  \dodoi{10.1103/PhysRevLett.119.161101}

\bibitem[{Abbott {et~al.}(2017{\natexlab{b}})}]{GW170817-MMA}
---. 2017{\natexlab{b}}, Astrophys. J., 848, L12,
  \dodoi{10.3847/2041-8213/aa91c9}

\bibitem[{{Abbott} {et~al.}(2017)}]{GW170817-HUBBLE}
{Abbott}, B.~P., {et~al.} 2017, \nat, 551, 85, \dodoi{10.1038/nature24471}

\bibitem[{Abbott {et~al.}(2017)}]{CE_PSD}
Abbott, B.~P., {et~al.} 2017, Classical and Quantum Gravity, 34, 044001,
  \dodoi{10.1088/1361-6382/aa51f4}

\bibitem[{Abbott {et~al.}(2018)}]{GW170817-TGR}
---. 2018.
\newblock \doarXiv{1811.00364}

\bibitem[{Abbott {et~al.}(2019{\natexlab{a}})}]{GW170817-SOURCE-PROPERTIES}
---. 2019{\natexlab{a}}, Physical Review X, 9, 011001,
  \dodoi{10.1103/PhysRevX.9.011001}

\bibitem[{Abbott {et~al.}(2019{\natexlab{b}})}]{GWTC-1}
---. 2019{\natexlab{b}}, Physical Review X, 9, 031040,
  \dodoi{10.1103/PhysRevX.9.031040}

\bibitem[{Abbott {et~al.}(2020{\natexlab{a}})}]{GW190412}
Abbott, R., {et~al.} 2020{\natexlab{a}}.
\newblock \doarXiv{2004.08342}

\bibitem[{Abbott {et~al.}(2020{\natexlab{b}})}]{GW190814}
---. 2020{\natexlab{b}}, The Astrophysical Journal, 896, L44,
  \dodoi{10.3847/2041-8213/ab960f}

\bibitem[{Adams {et~al.}(2016)Adams, Buskulic, Germain, Guidi, Marion, Montani,
  Mours, Piergiovanni, \& Wang}]{mbta}
Adams, T., Buskulic, D., Germain, V., {et~al.} 2016, Classical and Quantum
  Gravity, 33, 175012, \dodoi{10.1088/0264-9381/33/17/175012}

\bibitem[{Adhikari {et~al.}(2019)}]{Adhikari:2019zpy}
Adhikari, R.~X., {et~al.} 2019, Class. Quant. Grav., 36, 245010,
  \dodoi{10.1088/1361-6382/ab3cff}

\bibitem[{Bellm {et~al.}(2018)Bellm, Kulkarni, Graham, Dekany, Smith, Riddle,
  Masci, Helou, Prince, Adams, Barbarino, Barlow, Bauer, Beck, Belicki, Biswas,
  Blagorodnova, Bodewits, Bolin, Brinnel, Brooke, Bue, Bulla, Burruss, Cenko,
  Chang, Connolly, Coughlin, Cromer, Cunningham, De, Delacroix, Desai, Duev,
  Eadie, Farnham, Feeney, Feindt, Flynn, Franckowiak, Frederick, Fremling,
  Gal-Yam, Gezari, Giomi, Goldstein, Golkhou, Goobar, Groom, Hacopians, Hale,
  Henning, Ho, Hover, Howell, Hung, Huppenkothen, Imel, Ip, Ivezi{\'{c}},
  Jackson, Jones, Juric, Kasliwal, Kaspi, Kaye, Kelley, Kowalski, Kramer,
  Kupfer, Landry, Laher, Lee, Lin, Lin, Lunnan, Giomi, Mahabal, Mao, Miller,
  Monkewitz, Murphy, Ngeow, Nordin, Nugent, Ofek, Patterson, Penprase, Porter,
  Rauch, Rebbapragada, Reiley, Rigault, Rodriguez, van Roestel, Rusholme, van
  Santen, Schulze, Shupe, Singer, Soumagnac, Stein, Surace, Sollerman, Szkody,
  Taddia, Terek, Sistine, van Velzen, Vestrand, Walters, Ward, Ye, Yu, Yan, \&
  Zolkower}]{Bellm_2018}
Bellm, E.~C., Kulkarni, S.~R., Graham, M.~J., {et~al.} 2018, Publications of
  the Astronomical Society of the Pacific, 131, 018002,
  \dodoi{10.1088/1538-3873/aaecbe}

\bibitem[{Berger(2014)}]{Berger:2013jza}
Berger, E. 2014, Ann. Rev. Astron. Astrophys., 52, 43,
  \dodoi{10.1146/annurev-astro-081913-035926}

\bibitem[{{Burns} {et~al.}(2019)}]{FermiGBM}
{Burns}, E., {et~al.} 2019, \apj, 871, 90, \dodoi{10.3847/1538-4357/aaf726}

\bibitem[{{Cannon} {et~al.}(2012){Cannon}, {Cariou}, {Chapman},
  {Crispin-Ortuzar}, {Fotopoulos}, {Frei}, {Hanna}, {Kara}, {Keppel}, {Liao},
  {Privitera}, {Searle}, {Singer}, \& {Weinstein}}]{CannonEW}
{Cannon}, K., {Cariou}, R., {Chapman}, A., {et~al.} 2012, \apj, 748, 136,
  \dodoi{10.1088/0004-637X/748/2/136}

\bibitem[{Chu(2017)}]{spiir}
Chu, Q. 2017, PhD thesis, The University of Western Australia

\bibitem[{Connaughton {et~al.}(2016)Connaughton, Burns, Goldstein, Blackburn,
  Briggs, Zhang, Camp, Christensen, Hui, Jenke, Littenberg, McEnery, Racusin,
  Shawhan, Singer, Veitch, Wilson-Hodge, Bhat, Bissaldi, Cleveland,
  Fitzpatrick, Giles, Gibby, von Kienlin, Kippen, McBreen, Mailyan, Meegan,
  Paciesas, Preece, Roberts, Sparke, Stanbro, Toelge, \&
  Veres}]{Connaughton_2016}
Connaughton, V., Burns, E., Goldstein, A., {et~al.} 2016, The Astrophysical
  Journal, 826, L6, \dodoi{10.3847/2041-8205/826/1/l6}

\bibitem[{Coughlin {et~al.}(2018)Coughlin, Tao, Chan, Chatterjee, Christensen,
  Ghosh, Greco, Hu, Kapadia, Rana, Salafia, \& Stubbs}]{SlewingStrategy}
Coughlin, M.~W., Tao, D., Chan, M.~L., {et~al.} 2018, Monthly Notices of the
  Royal Astronomical Society, 478, 692, \dodoi{10.1093/mnras/sty1066}

\bibitem[{Cowperthwaite {et~al.}(2017)Cowperthwaite, Berger, Villar, Metzger,
  Nicholl, Chornock, Blanchard, Fong, Margutti, Soares-Santos, Alexander,
  Allam, Annis, Brout, Brown, Butler, Chen, Diehl, Doctor, Drout, Eftekhari,
  Farr, Finley, Foley, Frieman, Fryer, Garc{\'{i}}a-Bellido, {S Gill},
  Guillochon, Herner, Holz, Kasen, Kessler, Marriner, Matheson, Neilsen,
  Quataert, Palmese, Rest, Sako, Scolnic, Smith, Tucker, {G Williams},
  Balbinot, Carlin, Cook, Durret, Li, {A Lopes}, {C Louren{\c{c}}o}, Marshall,
  Medina, Muir, Mu{\~{n}}oz, Sauseda, Schlegel, Secco, Vivas, Wester, Zenteno,
  Zhang, {C Abbott}, Banerji, Bechtol, Benoit-L{\'{e}}vy, Bertin, Buckley-Geer,
  Burke, Capozzi, {Carnero Rosell}, Carrasco, Yanny, Zuntz, \&
  Woods}]{Cowperthwaite2017}
Cowperthwaite, P.~S., Berger, E., Villar, V.~A., {et~al.} 2017, ApJ, 848, L17,
  \dodoi{10.3847/2041-8213/aa8fc7}

\bibitem[{{Douchin} \& {Haensel}(2001)}]{SLy}
{Douchin}, F., \& {Haensel}, P. 2001, \aap, 380, 151,
  \dodoi{10.1051/0004-6361:20011402}

\bibitem[{{Drout} {et~al.}(2017){Drout}, {Piro}, {Shappee}, {Kilpatrick},
  {Simon}, {Contreras}, {Coulter}, {Foley}, {Siebert}, {Morrell}, {Boutsia},
  {Di Mille}, {Holoien}, {Kasen}, {Kollmeier}, {Madore}, {Monson},
  {Murguia-Berthier}, {Pan}, {Prochaska}, {Ramirez-Ruiz}, {Rest}, {Adams},
  {Alatalo}, {Ba{\~n}ados}, {Baughman}, {Beers}, {Bernstein}, {Bitsakis},
  {Campillay}, {Hansen}, {Higgs}, {Ji}, {Maravelias}, {Marshall}, {Moni Bidin},
  {Prieto}, {Rasmussen}, {Rojas-Bravo}, {Strom}, {Ulloa},
  {Vargas-Gonz{\'a}lez}, {Wan}, \& {Whitten}}]{Drout2017}
{Drout}, M.~R., {Piro}, A.~L., {Shappee}, B.~J., {et~al.} 2017, Science, 358,
  1570, \dodoi{10.1126/science.aaq0049}

\bibitem[{{Fairhurst}(2009)}]{Fairhurst1}
{Fairhurst}, S. 2009, New Journal of Physics, 11, 123006,
  \dodoi{10.1088/1367-2630/11/12/123006}

\bibitem[{{Fairhurst}(2011)}]{Fairhurst2}
---. 2011, Classical and Quantum Gravity, 28, 105021,
  \dodoi{10.1088/0264-9381/28/10/105021}

\bibitem[{Foucart(2012)}]{FoucartEMB}
Foucart, F. 2012, Phys. Rev. D, 86, 124007, \dodoi{10.1103/PhysRevD.86.124007}

\bibitem[{{Gehrels}(2004)}]{Swift}
{Gehrels}, N. 2004, in American Institute of Physics Conference Series, Vol.
  727, Gamma-Ray Bursts: 30 Years of Discovery, ed. E.~{Fenimore} \&
  M.~{Galassi}, 637--641, \dodoi{10.1063/1.1810924}

\bibitem[{Graham {et~al.}(2020)Graham, Ford, McKernan, Ross, Stern, Burdge,
  Coughlin, Djorgovski, Drake, Duev, Kasliwal, Mahabal, van Velzen, Belecki,
  Bellm, Burruss, Cenko, Cunningham, Helou, Kulkarni, Masci, Prince, Reiley,
  Rodriguez, Rusholme, Smith, \& Soumagnac}]{ZTF-BBH}
Graham, M.~J., Ford, K. E.~S., McKernan, B., {et~al.} 2020, Phys. Rev. Lett.,
  124, 251102, \dodoi{10.1103/PhysRevLett.124.251102}

\bibitem[{Harry {et~al.}(2018)Harry, Calderón~Bustillo, \&
  Nitz}]{Harry:2017weg}
Harry, I., Calderón~Bustillo, J., \& Nitz, A. 2018, Phys. Rev. D, 97, 023004,
  \dodoi{10.1103/PhysRevD.97.023004}

\bibitem[{Hild(2012)}]{ET_PSD}
Hild, S. 2012, Classical and Quantum Gravity, 29, 124006,
  \dodoi{10.1088/0264-9381/29/12/124006}

\bibitem[{{Hotokezaka} {et~al.}(2013){Hotokezaka}, {Kiuchi}, {Kyutoku},
  {Muranushi}, {Sekiguchi}, {Shibata}, \& {Taniguchi}}]{HotokezakaHNS}
{Hotokezaka}, K., {Kiuchi}, K., {Kyutoku}, K., {et~al.} 2013, \prd, 88, 044026,
  \dodoi{10.1103/PhysRevD.88.044026}

\bibitem[{Ivezi{\'{c}} {et~al.}(2019)Ivezi{\'{c}}, Kahn, Tyson, Abel, Acosta,
  Allsman, Alonso, AlSayyad, Anderson, Andrew, Angel, Angeli, Ansari,
  Antilogus, Araujo, Armstrong, Arndt, Astier, Aubourg, Auza, Axelrod, Bard,
  Barr, Barrau, Bartlett, Bauer, Bauman, Baumont, Bechtol, Bechtol, Becker,
  Becla, Beldica, Bellavia, Bianco, Biswas, Blanc, Blazek, Blandford, Bloom,
  Bogart, Bond, Booth, Borgland, Borne, Bosch, Boutigny, Brackett, Bradshaw,
  Brandt, Brown, Bullock, Burchat, Burke, Cagnoli, Calabrese, Callahan, Callen,
  Carlin, Carlson, Chandrasekharan, Charles-Emerson, Chesley, Cheu, Chiang,
  Chiang, Chirino, Chow, Ciardi, Claver, Cohen-Tanugi, Cockrum, Coles,
  Connolly, Cook, Cooray, Covey, Cribbs, Cui, Cutri, Daly, Daniel, Daruich,
  Daubard, Daues, Dawson, Delgado, Dellapenna, de~Peyster, de~Val-Borro, Digel,
  Doherty, Dubois, Dubois-Felsmann, Durech, Economou, Eifler, Eracleous,
  Emmons, Neto, Ferguson, Figueroa, Fisher-Levine, Focke, Foss, Frank, Freemon,
  Gangler, Gawiser, Geary, Gee, Geha, Gessner, Gibson, Gilmore, Glanzman,
  Glick, Goldina, Goldstein, Goodenow, Graham, Gressler, Gris, Guy, Guyonnet,
  Haller, Harris, Hascall, Haupt, Hernandez, Herrmann, Hileman, Hoblitt,
  Hodgson, Hogan, Howard, Huang, Huffer, Ingraham, Innes, Jacoby, Jain, Jammes,
  Jee, Jenness, Jernigan, Jevremovi{\'{c}}, Johns, Johnson, Johnson, Jones,
  Juramy-Gilles, Juri{\'{c}}, Kalirai, Kallivayalil, Kalmbach, Kantor, Karst,
  Kasliwal, Kelly, Kessler, Kinnison, Kirkby, Knox, Kotov, Krabbendam,
  Krughoff, Kub{\'{a}}nek, Kuczewski, Kulkarni, Ku, Kurita, Lage, Lambert,
  Lange, Langton, Guillou, Levine, Liang, Lim, Lintott, Long, Lopez, Lotz,
  Lupton, Lust, MacArthur, Mahabal, Mandelbaum, Markiewicz, Marsh, Marshall,
  Marshall, May, McKercher, McQueen, Meyers, Migliore, Miller, Mills, Miraval,
  Moeyens, Moolekamp, Monet, Moniez, Monkewitz, Montgomery, Morrison, Mueller,
  Muller, Arancibia, Neill, Newbry, Nief, Nomerotski, Nordby, O'Connor, Oliver,
  Olivier, Olsen, O'Mullane, Ortiz, Osier, Owen, Pain, Palecek, Parejko,
  Parsons, Pease, Peterson, Peterson, Petravick, Petrick, Petry, Pierfederici,
  Pietrowicz, Pike, Pinto, Plante, Plate, Plutchak, Price, Prouza, Radeka,
  Rajagopal, Rasmussen, Regnault, Reil, Reiss, Reuter, Ridgway, Riot, Ritz,
  Robinson, Roby, Roodman, Rosing, Roucelle, Rumore, Russo, Saha, Sassolas,
  Schalk, Schellart, Schindler, Schmidt, Schneider, Schneider, Schoening,
  Schumacher, Schwamb, Sebag, Selvy, Sembroski, Seppala, Serio, Serrano, Shaw,
  Shipsey, Sick, Silvestri, Slater, Smith, Smith, Sobhani, Soldahl,
  Storrie-Lombardi, Stover, Strauss, Street, Stubbs, Sullivan, Sweeney,
  Swinbank, Szalay, Takacs, Tether, Thaler, Thayer, Thomas, Thornton, Thukral,
  Tice, Trilling, Turri, Berg, Berk, Vetter, Virieux, Vucina, Wahl, Walkowicz,
  Walsh, Walter, Wang, Wang, Warner, Wiecha, Willman, Winters, Wittman, Wolff,
  Wood-Vasey, Wu, Xin, Yoachim, \& Zhan}]{Ivezic_2019}
Ivezi{\'{c}}, {\v{Z}}., Kahn, S.~M., Tyson, J.~A., {et~al.} 2019, The
  Astrophysical Journal, 873, 111, \dodoi{10.3847/1538-4357/ab042c}

\bibitem[{{KAGRA Collaboration} {et~al.}(2019){KAGRA Collaboration}, {LIGO
  Scientific Collaboration}, \& {Virgo Collaboration}}]{observer_summary}
{KAGRA Collaboration}, {LIGO Scientific Collaboration}, \& {Virgo
  Collaboration}. 2019, Advanced LIGO, Advanced Virgo and KAGRA observing run
  plans.
\newblock
  \url{https://dcc.ligo.org/public/0161/P1900218/002/SummaryForObservers.pdf}

\bibitem[{{Kasen} {et~al.}(2017){Kasen}, {Metzger}, {Barnes}, {Quataert}, \&
  {Ramirez-Ruiz}}]{GW170817-HEAVY-ELEMENTS}
{Kasen}, D., {Metzger}, B., {Barnes}, J., {Quataert}, E., \& {Ramirez-Ruiz}, E.
  2017, \nat, 551, 80, \dodoi{10.1038/nature24453}

\bibitem[{Kyutoku {et~al.}(2010)Kyutoku, Shibata, \& Taniguchi}]{2H}
Kyutoku, K., Shibata, M., \& Taniguchi, K. 2010, Phys. Rev. D, 82, 044049,
  \dodoi{10.1103/PhysRevD.82.044049}

\bibitem[{{LIGO Scientific Collaboration}(2015)}]{Voyager_PSD}
{LIGO Scientific Collaboration}. 2015, Instrument Science White Paper.
\newblock \url{https://dcc.ligo.org/public/0120/T1500290/002/T1500290.pdf}

\bibitem[{{LIGO Scientific Collaboration}(2020)}]{lalsuite}
---. 2020, {LIGO} {A}lgorithm {L}ibrary - {LALS}uite, free software (GPL),
  \dodoi{10.7935/GT1W-FZ16}

\bibitem[{Loeb(2016)}]{Loeb_2016}
Loeb, A. 2016, The Astrophysical Journal, 819, L21,
  \dodoi{10.3847/2041-8205/819/2/l21}

\bibitem[{London {et~al.}(2018)London, Khan, Fauchon-Jones, Garc\'{\i}a,
  Hannam, Husa, Jim\'enez-Forteza, Kalaghatgi, Ohme, \&
  Pannarale}]{imrphenomhm}
London, L., Khan, S., Fauchon-Jones, E., {et~al.} 2018, Phys. Rev. Lett., 120,
  161102, \dodoi{10.1103/PhysRevLett.120.161102}

\bibitem[{Messick {et~al.}(2017)Messick, Blackburn, Brady, Brockill, Cannon,
  Cariou, Caudill, Chamberlin, Creighton, Everett, Hanna, Keppel, Lang, Li,
  Meacher, Nielsen, Pankow, Privitera, Qi, Sachdev, Sadeghian, Singer, Thomas,
  Wade, Wade, Weinstein, \& Wiesner}]{gstlal}
Messick, C., Blackburn, K., Brady, P., {et~al.} 2017, Phys. Rev. D, 95, 042001,
  \dodoi{10.1103/PhysRevD.95.042001}

\bibitem[{{Metzger}(2017)}]{MetzgerKN}
{Metzger}, B.~D. 2017, Living Reviews in Relativity, 20, 3,
  \dodoi{10.1007/s41114-017-0006-z}

\bibitem[{{Nakar}(2007)}]{NakarGRB}
{Nakar}, E. 2007, \physrep, 442, 166, \dodoi{10.1016/j.physrep.2007.02.005}

\bibitem[{Newman \& Penrose(1966)}]{NewmanPenrose}
Newman, E.~T., \& Penrose, R. 1966, Journal of Mathematical Physics, 7, 863,
  \dodoi{10.1063/1.1931221}

\bibitem[{Nitz {et~al.}(2020)Nitz, Harry, Brown, Biwer, Willis, Canton, Capano,
  Pekowsky, Dent, Williamson, Davies, De, Cabero, Machenschalk, Kumar, Reyes,
  Macleod, Pannarale, dfinstad, Massinger, Tápai, Singer, Khan, Fairhurst,
  Kumar, Nielsen, shasvath, Dorrington, Lenon, \& Gabbard}]{pycbc_software}
Nitz, A., Harry, I., Brown, D., {et~al.} 2020, gwastro/pycbc: PyCBC Release
  1.16.4, v1.16.4,  Zenodo, \dodoi{10.5281/zenodo.3904502}

\bibitem[{Nitz {et~al.}(2018)Nitz, Dal~Canton, Davis, \& Reyes}]{pycbc}
Nitz, A.~H., Dal~Canton, T., Davis, D., \& Reyes, S. 2018, Phys. Rev. D, 98,
  024050, \dodoi{10.1103/PhysRevD.98.024050}

\bibitem[{Punturo {et~al.}(2010)Punturo, Abernathy, Acernese, Allen, Andersson,
  Arun, Barone, Barr, Barsuglia, Beker, Beveridge, Birindelli, Bose, Bosi,
  Braccini, Bradaschia, Bulik, Calloni, Cella, Mottin, Chelkowski, Chincarini,
  Clark, Coccia, Colacino, Colas, Cumming, Cunningham, Cuoco, Danilishin,
  Danzmann, Luca, Salvo, Dent, Rosa, Fiore, Virgilio, Doets, Fafone, Falferi,
  Flaminio, Franc, Frasconi, Freise, Fulda, Gair, Gemme, Gennai, Giazotto,
  Glampedakis, Granata, Grote, Guidi, Hammond, Hannam, Harms, Heinert, Hendry,
  Heng, Hennes, Hild, Hough, Husa, Huttner, Jones, Khalili, Kokeyama, Kokkotas,
  Krishnan, Lorenzini, Lück, Majorana, Mandel, Mandic, Martin, Michel,
  Minenkov, Morgado, Mosca, Mours, Müller{\textendash}Ebhardt, Murray,
  Nawrodt, Nelson, Oshaughnessy, Ott, Palomba, Paoli, Parguez, Pasqualetti,
  Passaquieti, Passuello, Pinard, Poggiani, Popolizio, Prato, Puppo, Rabeling,
  Rapagnani, Read, Regimbau, Rehbein, Reid, Rezzolla, Ricci, Richard, Rocchi,
  Rowan, Rüdiger, Sassolas, Sathyaprakash, Schnabel, Schwarz, Seidel, Sintes,
  Somiya, Speirits, Strain, Strigin, Sutton, Tarabrin, Thüring, van~den Brand,
  van Leewen, van Veggel, van~den Broeck, Vecchio, Veitch, Vetrano, Vicere,
  Vyatchanin, Willke, Woan, Wolfango, \& Yamamoto}]{ET}
Punturo, M., Abernathy, M., Acernese, F., {et~al.} 2010, Classical and Quantum
  Gravity, 27, 194002, \dodoi{10.1088/0264-9381/27/19/194002}

\bibitem[{{Reitze} {et~al.}(2019){Reitze}, {Adhikari}, {Ballmer}, {Barish},
  {Barsotti}, {Billingsley}, {Brown}, {Chen}, {Coyne}, {Eisenstein}, {Evans},
  {Fritschel}, {Hall}, {Lazzarini}, {Lovelace}, {Read}, {Sathyaprakash},
  {Shoemaker}, {Smith}, {Torrie}, {Vitale}, {Weiss}, {Wipf}, \& {Zucker}}]{CE}
{Reitze}, D., {Adhikari}, R.~X., {Ballmer}, S., {et~al.} 2019, in \baas,
  Vol.~51, 35.
\newblock \doarXiv{1907.04833}

\bibitem[{Sathyaprakash(1994)}]{Sathyaprakash:1994nj}
Sathyaprakash, B. 1994, Phys. Rev. D, 50, 7111,
  \dodoi{10.1103/PhysRevD.50.R7111}

\bibitem[{Sesana(2016)}]{SesanaMultiband}
Sesana, A. 2016, Phys. Rev. Lett., 116, 231102,
  \dodoi{10.1103/PhysRevLett.116.231102}

\bibitem[{{Tsang} {et~al.}(2012){Tsang}, {Read}, {Hinderer}, {Piro}, \&
  {Bondarescu}}]{2012PhRvL.108a1102T}
{Tsang}, D., {Read}, J.~S., {Hinderer}, T., {Piro}, A.~L., \& {Bondarescu}, R.
  2012, \prl, 108, 011102, \dodoi{10.1103/PhysRevLett.108.011102}

\bibitem[{Unnikrishnan(2013)}]{LIGO-INDIA}
Unnikrishnan, C.~S. 2013, International Journal of Modern Physics D, 22,
  1341010, \dodoi{10.1142/s0218271813410101}

\bibitem[{Varma {et~al.}(2014)Varma, Ajith, Husa, Bustillo, Hannam, \&
  Pürrer}]{Varma:2014jxa}
Varma, V., Ajith, P., Husa, S., {et~al.} 2014, Phys. Rev. D, 90, 124004,
  \dodoi{10.1103/PhysRevD.90.124004}

\end{thebibliography}

\end{document}